**Quantum imaging: Scattered observations on "Copenhagen"**

**Adrian Kent, Centre for Quantum Computation, Department of Applied Mathematics and Theoretical Physics, University of Cambridge, Cambridge, U.K.** *and* **Perimeter Institute for Theoretical Physics, Waterloo, Ontario, Canada.**

Having long been fascinated by Michael Frayn's work, I could not resist the editors' kind invitation to offer some impressions of his play, Copenhagen, from the perspective of a quantum physicist. As well as the the original 1998 National Theatre production and the play's text, these are influenced by Frayn's intriguing metaphysical books Constructions (1974) and The Human Touch (2006), which explore more directly some of the questions that the play raises about human narratives, knowledge and motivations, and their possible connection to fundamental physics.

The extent to which Copenhagen respects historical fact has been much discussed. Not being competent to judge, I have to admit that my appreciation of the play relies on a less stringent test: that Frayn's characters, their reconstructed conversations and their conjectured motives should have some interesting correlations -- a quantum physicist would say, a significant inner product -- with historical and psychological reality. By this criterion, I think the play succeeds superbly: Frayn gets the scientific technicalities right, captures brilliantly physicists' ways of thinking and conversational style, and gives us believable versions of Heisenberg and of Niels and Magrethe Bohr.

Frayn tells his story as a series of ``drafts'', hypothetical and seemingly incompatible histories. Within and around these drafts are many images and metaphors drawn from quantum physics -- in particular Bohr's principle of complementarity and Heisenberg's uncertainty principle. These are certainly enjoyably playful analogies, which hint at speculations of a deeper connection, and could even be read as insisting that science teaches us that elementary particles and human beings alike necessarily resist any more precise analysis.

However, this last reading, I think, should itself be resisted. Postmodernists and historical relativists have indeed, of course, often called on quantum theory in their support. And indeed, to be fair, one of the deepest and most beautiful ways of formulating quantum theory describes the evolution of states over time as the collective result of a sum over alternative histories. Indeed, too, defining and justifying the concept of an unambiguous past remains a deep unsolved problem in modern quantum cosmology. On the other hand, physicists tend to wonder how it can possibly be consistent to rely on -- and, most of us would say, wildly extrapolate -- features of one scientific theory in order to argue for a radically anti-scientific world view.

In any case, Frayn goes deeper. One of the most intriguing parts of the play, for me, was Magrethe's argument that the Copenhagen interpretation of quantum theory, while purporting to subsume complementarity and uncertainty within a complete and consistent metaphysical account of science, should be understood as something more in the nature of a political treaty than a hard-won intellectual revelation arrived at purely by logical analysis of experimental data.

Magrethe's target is not quantum theory itself, but the philosophical clothing in which many of its founders dressed it. They were led empirically by a series of quite unexpected experimental results, and created a quite extraordinarily successful theory -- one that predicts mathematically the outcomes of experiments and observations covering a vast range of physical phenomena, and to very high precision. And yet, uncomfortable though it is for many physicists to accept, the theoretical physics community is of course not exempt from the natural human tendencies towards herding and constructing social hierarchies, which tend to develop consensus views by extra-scientific means.

It is the tension between these two dynamics that makes the intellectual history of quantum theory so sociologically fascinating. It explains how it can nowadays be respectable for a physicist to say that she does not really know what precisely Bohr and his followers meant by the Copenhagen interpretation or the principle of complementarity, and to wonder how these cloudy and at best provisional ideas so dominated physicists' understanding of quantum theory for so long -- a view which was close to heresy between around 1930 and 1980. The fact is, if one is thoughtful and open-minded, when one tries to extract an account of the ultimate natures of reality or scientific knowledge from quantum theory, one finds oneself once again surrounded by drafts. There is ample motivation to tweak an equation here, or vary an axiom there, or take a new physical perspective -- changes whose empirical consequences are imperceptible, but which produce radically different resulting stories about reality and our experience. For now -- until we find better ideas or a better theory -- scientists and artists alike can indulge their tastes, picking and choosing from a startlingly wide range of metaphysical views that seem consistent with quantum theory. One of the beauties of the play, for me, is its openness to this ambiguity, subtly querying its own central metaphors at the same time as using them to query other claims to truth.

Could there yet turn out to be some firm scientific basis to these metaphors, though? Could we, for example, ultimately find a deep connection between quantum theory and human consciousness? This question intrigues many physicists. To my mind, two of the most fascinating problems we face in understanding nature are the problem of finding a mathematical description of reality consistent with quantum theory, and of finding a fundamental physical theory of consciousness. Perhaps the questions are ill-posed, but one can at least imagine possible ways in which they could have elegant scientific solutions, and -- albeit yet more vaguely, speculatively and uncompellingly -- possible forms of mathematical law connecting the two. So I don't think the question is meaningless or will necessarily ultimately prove fruitless. However, we don't seem to have even the beginnings of a physical theory of consciousness: if there actually is such a thing, we may be further from uncovering it than the ancient Greeks were from modern physics. All in all, I'm not too optimistic that we will make any scientific discoveries relating consciousness and quantum theory any time soon.

I thought I would try to conclude on a less analytic and more playful note, by seeing if I could suggest a new parallel between physics and human behaviour that might appeal to Frayn's readers. In Copenhagen, human beings, under Frayn's microscope, are seen as something more like miniature communities, with their own internal politics, power struggles, and internal compromises -- an analogy made more explicit in "The Human Touch" and his later play, "Democracy". To a physicist, this brings to mind the self-similar scaling laws governing many statistical systems near phase transitions and fractals, whose properties, viewed at any two fixed length scales, are intimately mathematically related --- but not generally identical. For example, the strength of the interaction between neighbouring elements in a statistical system may depend on the ratio of the scales. One could try to improve Frayn's model by thinking of ourselves as some sort of mathematically rescaled community. But then which features should be rescaled? Could this model of a self containing Whitmanesque multitudes (which contain their own multitudes, and so on) perhaps even guide us to interestingly different ways of weighting the inputs of different parts of our selves when arriving at life decisions? For example, there are solid arguments in favour of giving regions (or other subcomponents) in a nation a representation in the government proportional to the square root of their population [1]. Political complexities may generally prevent this mathematical ideal being realised in the government of nations, but there is nothing to stop us (if we really think we can make sense of the idea) from implementing this or other interesting and novel strategies when governing our own behaviour. In fact, anthropologists and social scientists have already given much thought [2] to the idea of a parallel between mathematical self-similarity and human communities. So maybe there is limited new mileage in these admittedly undeveloped thoughts. Still, I thought I would offer them, in a spirit of cross-disciplinary experiment, in case some readers might possibly find some creative uses for them.

[1] See for example L.S.Penrose, The elementary statistics of majority voting, Journal of the Royal Statistical Society 109, pp 53-57.
[2] E.g. On The Order of Chaos: Social Anthropology and the Science of Chaos, M.S. Mosko and F.H. Damon (eds.), (Berghahn Books, 2005).